\documentstyle[aps,prb,epsfig,fancyhdr,multicol]{revtex}

\begin{document}



\title{MIKA: a multigrid-based program package for electronic structure calculations}

\author{T.~Torsti, M. Heiskanen, M.~J.~Puska, and R.~M.~Nieminen}

\address{Laboratory of Physics, Helsinki University of Technology,
P.O. Box 1100, FIN-02015 HUT, FINLAND}

\date{\today}
\maketitle

\thispagestyle{fancy}

\begin{abstract}
A general real-space multigrid algorithm MIKA (Multigrid Instead
of the K-spAce) for the self-consistent solution
of the Kohn-Sham equations appearing in the state-of-the-art electronic-structure
calculations is described. The most important part
of the method is the multigrid solver for the Schr\"odinger equation.
Our choice is the Rayleigh quotient multigrid method (RQMG), which applies
directly to the minimization of the Rayleigh quotient on the finest level. Very
coarse correction grids can be used, because there is in principle
no need  to represent the states on the coarse levels. The RQMG method is
generalized for the simultaneous solution of all the states of the system
using a penalty functional to keep the states orthogonal. Special care
has been taken to optimize the iterations towards  the self-consistency
and to run the code in parallel computer architectures. 
The scheme has been implemented in multiple geometries. We show examples from
electronic structure calculations employing nonlocal pseudopotentials
and/or the jellium model. The RQMG solver is also applied for the calculation
of positron states in solids.
\end{abstract}

\begin{multicols}{2}
\narrowtext

\section{Introduction}
\label{sec:introduction}

The goal of computational materials science and also that of modeling
of nanoscale man-made structures is to calculate from first principles 
the various chemical and/or physical properties. This requires the solution 
of the electronic (and ionic) structures of the system in question. The 
density-functional theory (DFT) \cite{kohn98} makes a huge step towards this 
goal by casting the untractable problem of many interacting electrons
to that of noninteracting particles under the influence of an effective
potential. 
However, in order to apply DFT in practice one has to resort to
approximations for electron exchange and correlation such as the
local-density approximation (LDA) or the generalized-gradient
approximation (GGA). Moreover, in the case of systems
consisting of hundreds or more atoms it is still a challenge
to solve numerically efficiently for the ensuing Kohn-Sham
equations.

We have developed a real-space multigrid method called MIKA (Multigrid Instead
of the K-spAce) for the numerical solution of the
Kohn-Sham equations \cite{mgarticle1}. In real-space 
methods\cite{beckrev,arias,waghmare}, the values of the
wave-functions and potentials are presented using three-dimensional point
grids, and the partial differential equations are discretized using finite
differences. Multigrid methods\cite{brandt1,beckrev} overcome
the critical slowing-down (CSD) phenomenon occuring with basic real-space relaxation methods.
Several approaches employing the multigrid idea have appeared during recent
years\cite{briggs,ancilotto,fattebert2,wang1}.

From the different multigrid methods available for the solution of the Schr\"odinger
equation, we have picked up the Rayleigh Quotient Multigrid (RQMG) method introduced
by Mandel and McCormick \cite{McCormick}. This approach differs from
full-approximation-storage\cite{brandt2,beck1,wang1,costiner} (FAS) methods, as well
as from those methods\cite{briggs}, where the eigenproblem is linearized.

In the RQMG method the coarse grid relaxation passes
are performed so that the Rayleigh quotient calculated on the {\em fine}
grid will be minimized. In this way there is no requirement for the
solution to be well represented on a coarse grid and the coarse grid
representation problem is avoided. Mandel and McCormick\cite{McCormick}
introduced the method for the solution of the eigenpair corresponding to
the lowest eigenvalue. We have generalized it to the simultaneous
solution of a desired number of lowest eigenenergy states by developing
a scheme which keeps the eigenstates separated by the use of a
penalty functional\cite{mgarticle1}. 

\section{Numerical Methods}
\label{sec:methods}

In our RQMG application the coarse grid relaxations are performed by the 
so-called coordinate relaxation method. It solves the discretized
eigenproblem
\begin{equation}
  H u = \lambda B u
\end{equation}
by minimizing the Rayleigh quotient
\begin{equation}
\label{Ray}
  \frac{\langle u\arrowvert H\arrowvert u\rangle}
       {\langle u\arrowvert B\arrowvert u\rangle}.
\end{equation}
Above, $H$ and $B$ are matrix operators chosen so that the Schr\"odinger
equation discretized on a real-space point grid with spacing $h$
is satisfied to a chosen order $O(h^n)$.
In Eq. (\ref{Ray}) $u$ is a vector containing
the wave function values at the grid points. In the relaxation
method, the current estimate $u$ is replaced by $ u' = u + \alpha d$,
where the search vector $d$ is simply chosen to be unity in one grid point
and to vanish in all other points, and $\alpha$ is chosen to minimize the
Rayleigh quotient. This leads to a simple \footnote{For the sake of simplicity, 
the wave-functions, and thus $\alpha$, are here assumed real. We have 
implemented the complex case as well.} quadratic equation for $\alpha$.
A complete coordinate relaxation
pass is then obtained by performing the minimization at each point
in turn and these passes can be repeated until the lowest state is
found with desired accuracy.

\begin{figure}[h]
\centerline{\resizebox{\columnwidth}{!}{\includegraphics{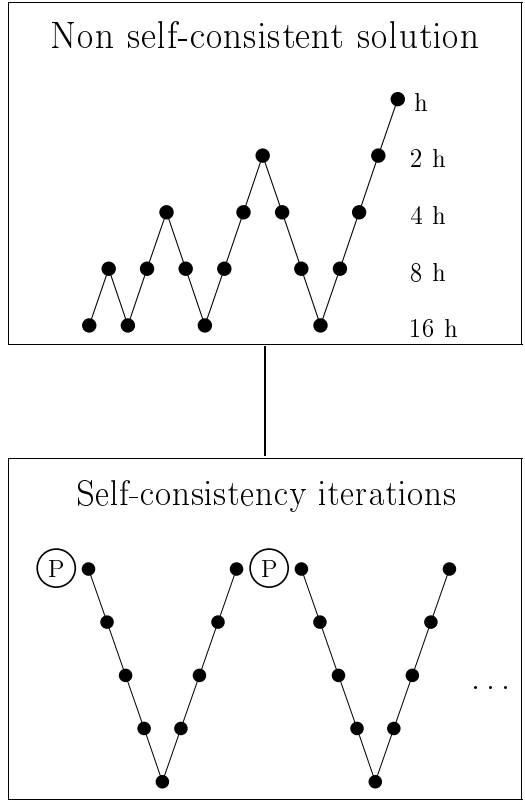}}}
\vspace{0.5cm}
\caption{Strategy of self-consistency iterations in MIKA. First, the wavefunctions are solved 
nonselfconsistently using the full multigrid method in the initial
potential corresponding to the superposition of pseudoatoms.
Then the effective potential is updated (this is denoted by P in the figure).
The potential update amounts to calculation of the new electron density,
the solution of the Poisson equation and calculation of the new exchange correlation
potential. Next the wave-functions are updated by one V-cycle. These two steps are repeated
until self-consistency has been reached.}
\label{fig:strategy}
\end{figure}

Naturally, also the coordinate relaxation suffers from CSD
because of the use of local information only in
updating $u$ in a certain point. In order to avoid it one applies the
multigrid idea. In the multigrid scheme by Mandel and
McCormick\cite{McCormick}
the crucial point is that {\em coarse} grid coordinate relaxation passes
are performed so that the Rayleigh quotient calculated on the {\em fine}
grid will be minimized. In this way there is no requirement for the
solution to be well represented on a coarse grid. In practice, a
coarse grid search substitutes the fine grid solution by
\begin{equation}
\label{rqmgchgeq}
 u_f' = u_f + \alpha I_c^f e_c,
\end{equation}
where the subscripts $f$ and $c$ stand for the fine and coarse
grids, respectively, and $I_c^f$ a prolongation operator interpolating
the coarse grid vector to the fine grid. The Rayleigh quotient to
be minimized is then
\begin{eqnarray}
\label{rqmgeq}
&    \frac{\langle u_f + \alpha I_c^f d_c \arrowvert H_f \arrowvert
         u_f + \alpha I_c^f d_c \rangle}
       {\langle u_f + \alpha I_c^f d_c \arrowvert B_f \arrowvert
         u_f + \alpha I_c^f d_c \rangle} = \qquad \qquad \qquad \qquad \nonumber \\
&\qquad \qquad \qquad   
      \frac{ \langle u_f \arrowvert H_f u_f \rangle
         + 2\alpha \langle I_f^c H_f u_f \arrowvert d_c \rangle
         + \alpha^2 \langle d_c \arrowvert H_c d_c \rangle
       }
       { \langle u_f \arrowvert B_f u_f \rangle
         + 2\alpha \langle I_f^c B_f u_f \arrowvert d_c \rangle
         + \alpha^2 \langle d_c \arrowvert B_c d_c \rangle
       }.
\end{eqnarray}
The second form is obtained by relating the coarse grid operators,
$H_c$ and $B_c$, with the fine grid ones, $H_f$ and $B_f$, by the
Galerkin condition
\begin{equation}
\label{galerkincond}
   H_c  = I_f^c H_f I_c^f; \quad
   B_c  = I_f^c B_f I_c^f; \quad
   I_f^c = \left(I_c^f\right)^T.
\end{equation}
The key point to note is that when $H_f u_f$ and $B_f u_f$ are provided
from the fine grid to the coarse grid, the remaining
integrals can be calculated on the coarse grid itself. Thus one really
applies coordinate relaxation on the coarse grids to minimize the
\textit{fine level} Rayleigh quotient. This is a major departure from
the earlier methods, which to some extent rely on the ability to
represent the solution of some coarse grid equation on the coarse grid
itself. Here, on the other hand, one can calculate the \textit{exact}
change in the Rayleigh quotient due to \textit{any} coarse grid
change, no matter how coarse the grid itself is. There is no equation
whose solution would have to be representable.

In the MIKA package we have generalized the RQMG method to the simultaneous
solution of several mutually orthogonal eigenpairs. The separation
of the different states is divided into two or three subtasks. First,
in order to make the coarse grid relaxations converge towards the
desired state we apply a penalty functional scheme.
Given the current approximations for the $k$ lowest eigenfunctions,
the next lowest, $(k+1)$'th  state
is updated by minimizing the functional
\begin{equation}
\label{rqmgneq}
 \frac{\langle u_{k+1}\arrowvert H\arrowvert u_{k+1}\rangle}
      {\langle u_{k+1}\arrowvert B\arrowvert u_{k+1}\rangle}
 + \sum\limits_{i=1}^{k}
    q_i \frac{\left|\langle u_i | u_{k+1}\rangle\right|^2}
             {\langle u_i | u_i\rangle \cdot
              \langle u_{k+1} | u_{k+1}\rangle}.
\end{equation}
The minimization of this functional is equivalent to imposing
the orthonormality constraints against the lower $k$ states,
when $q_i \rightarrow \infty$. By increasing the shifts $q_i$
any desired accuracy can be obtained, but in order to obtain
a computationally efficient algorithm a reasonable finite value should
be used, for example
\begin{equation}
  q_i = (\lambda_{k+1}-\lambda_i) + {\rm Q},
\end{equation}
where $Q$ is a sufficiently large positive constant. In our test
calculations $Q$ is of the order of $Q=0.5\ldots 2$ Ha.

\begin{figure}[h]
\centerline{\framebox{\resizebox{0.7\columnwidth}{!}{\includegraphics{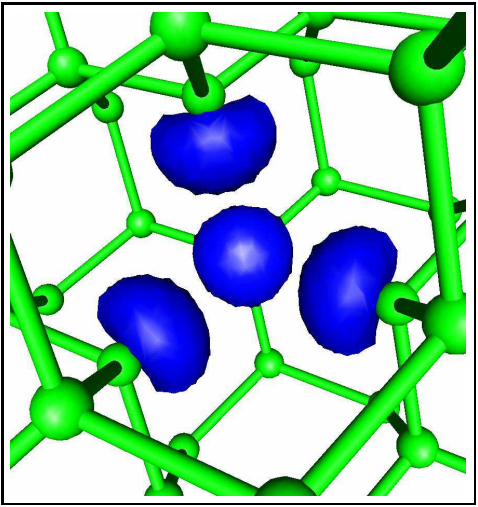}}}}
\vspace{0.5cm}
\caption{Electron density isosurface of the deep state localized at a 
neutral, ideal (no ion relaxation) vacancy in bulk Si.}
\label{fig:deep_state}
\end{figure}

The substitution (\ref{rqmgchgeq}) is introduced in the functional
~(\ref{rqmgneq}) and the minimization with respect to $\alpha$
leads again to a quadratic equation. This time the coefficients
contain terms due to the penalty part.

While the penalty functional keeps the states separated on the coarse levels,
we apply a simple relaxation method (Gauss-Seidel) on the finest level.
The Gauss-Seidel method converges to the nearest eigenvalue, so ideally
no additional orthogonalizations would be needed. In practice, however,
we use Gramm-Schmidt orthogonalizations and subspace rotations\cite{mgarticle1}.
However, the number of fine grid orthogonalizations remains quite
plausible, for example, in comparison with the conjugate gradient
search of eigenpairs employing only the finest grid \cite{seitsonen}.

The Kohn-Sham equations have to be solved self-consistently, {\em i.e.}
the wave functions solved from the single-particle equation
determine via the density (solution of the Poisson equation
and the calculation of the exchange-correlation potential)
the effective potential for which they
should again be solved. To approach this self-consistency
requires an optimized strategy so that numerical accuracy of the wave
functions and the potential increase in balance, enabling the most
efficient convergence \cite{wang1,waghmare}. Our strategy in MIKA 
for self-consistency iterations is illustrated in Fig. \ref{fig:strategy}. 
The Poisson equation for the Coulomb potential is solved also by the 
multigrid method.

\begin{figure}[h]
\centerline{{\resizebox{0.5\columnwidth}{!}{\includegraphics{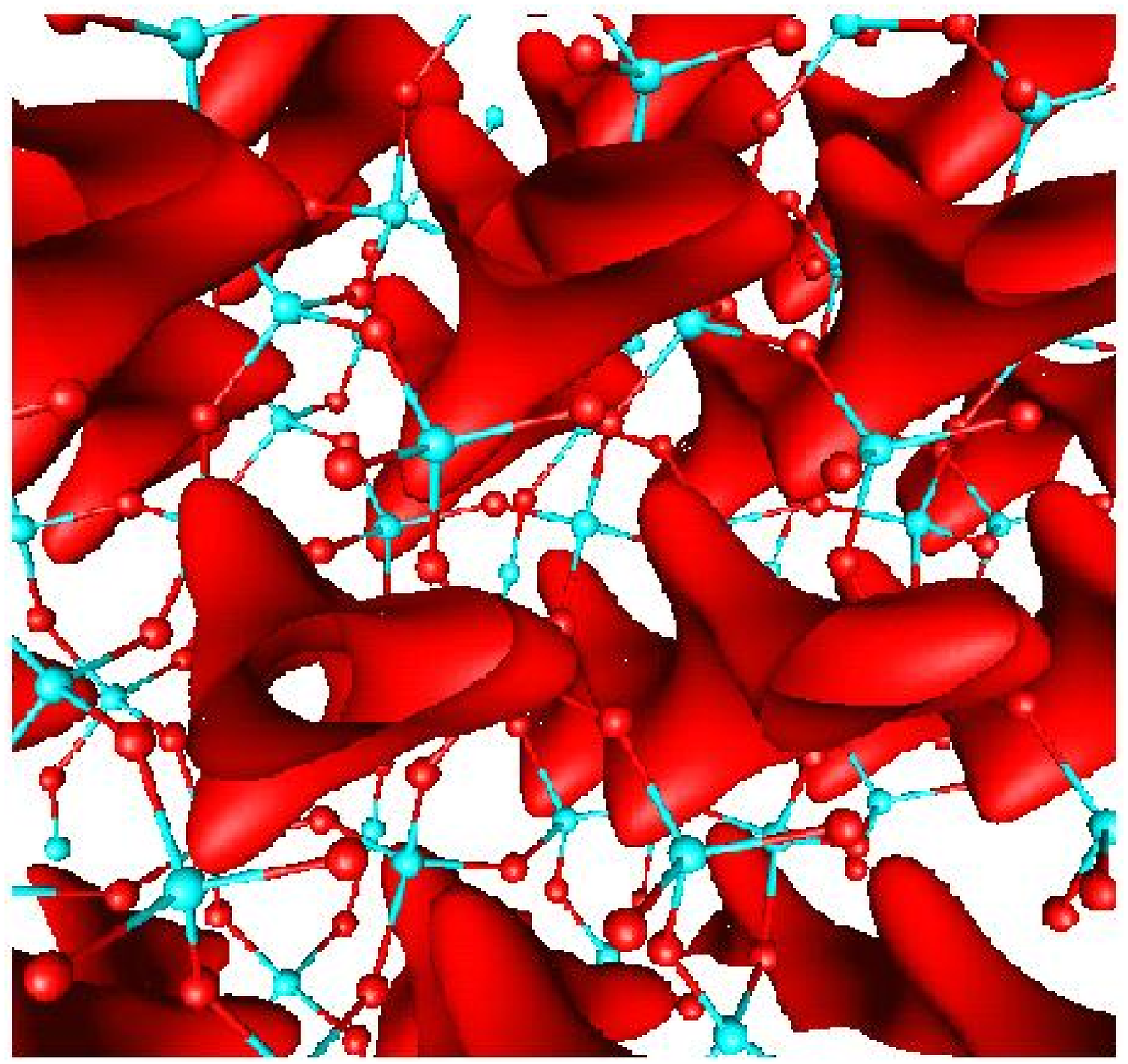}}\resizebox{0.5\columnwidth}{!}{\includegraphics{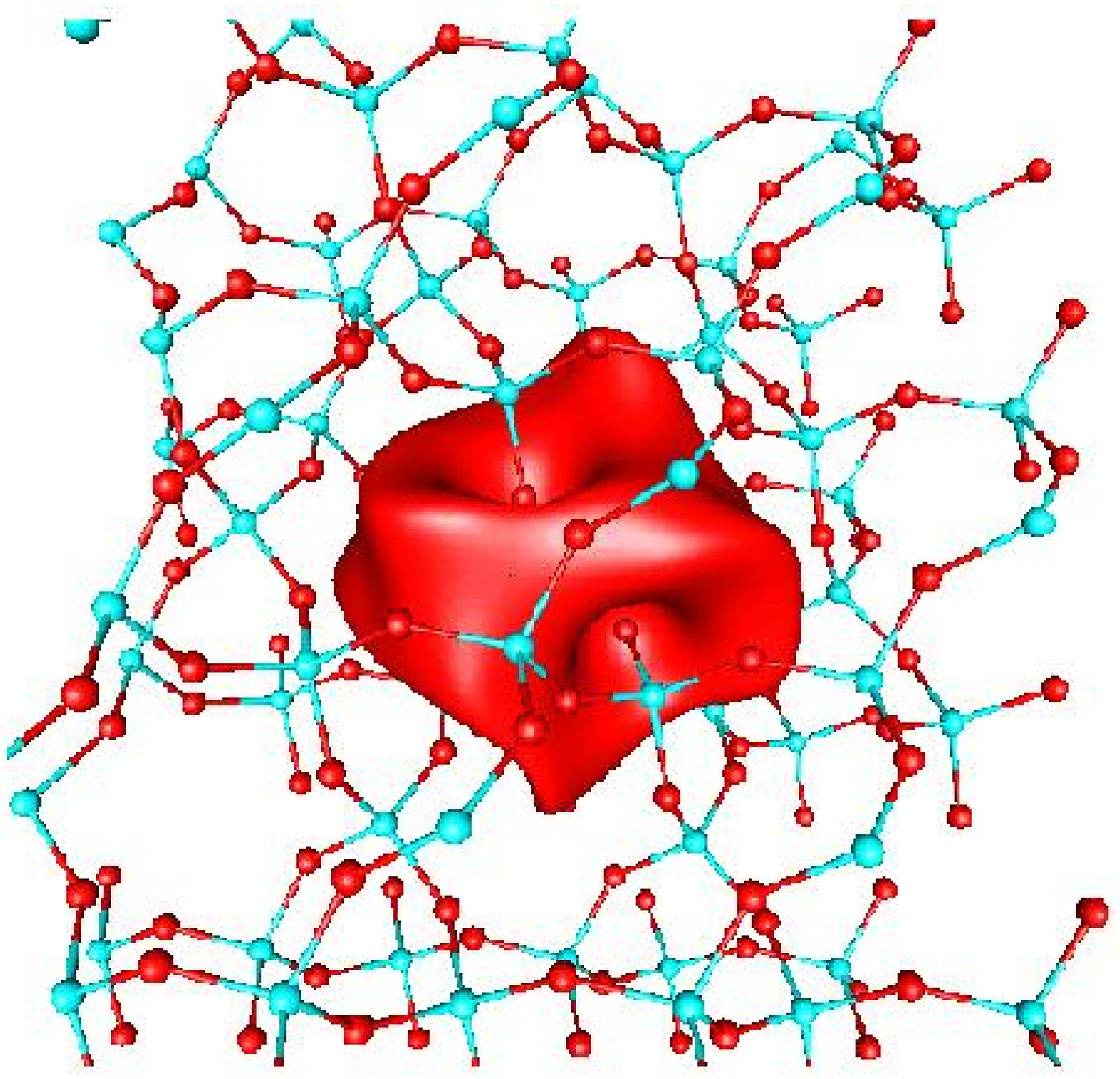}}}}
\vspace{0.5cm}
\caption{Left panel: Isosurface of the delocalized positron state in the perfect bulk SiO$_2$. 
Right panel: Isosurface of the positron state localized at a Si vacancy in SiO$_2$.}
\label{fig:positrons}
\end{figure}

\section{Examples}
\label{sec:results}

We have demonstrated\cite{mgarticle1} the performance of the MIKA scheme 
in calculating the electronic structures of small molecules
and solid-state systems described by pseudopotentials.
As a typical application Fig. \ref{fig:deep_state} shows
the electron density of the so-called deep state localized at a
neutral, ideal vacancy in bulk Si.
It was shown, that the accuracy of 1 meV for the total energy was reached
after three or four V-cycles, and that the amount of cpu-time needed
was of the same order as when applying state-of-the-art plane-wave codes.
We obtained an average convergence rate  of
approximately one decade per self-consistency iteration. This is of the same
order as those reported by Wang and Beck \cite{wang1} in their FAS scheme
or by Kresse and Furthm\"uller \cite{kresse2} in their plane-wave scheme
employing self-consistency iterations. The convergence rate of one decade per
self-consistency iteration is better than that obtained by Ancilotto
{\em et al.} \cite{ancilotto} in the FMG scheme and much better than
the rate reached in the linearized multigrid scheme by Briggs {\em et al.}
\cite{briggs}.

We have applied the RQMG method also for the calculation of positron
states in solids. Fig. \ref{fig:positrons} shows how the delocalized
positron state in the perfect $\alpha$-quartz is trapped in to a
Si-vacancy. Positron states are a particularly simple case for our method, because
only the lowest energy wave function needs to be calculated in a given potential,
so that no orthogonalizations or penalty functionals are needed. Moreover,
in a simple scheme an electron density calculated without the influence of
the positron can be used as the starting point \cite{pos_rev}.
However, even for the positron states the superior performance of the multigrid 
method in comparison with straightforward relaxation schemes is evident. 
For example, we have calculated the
positron state in the Si-vacancy in bulk Si using a supercell containing 1727
atoms. The solution of the positron wave-function using the RQMG-method
took less than a minute of cpu-time on a typical work station.
To put this in proper context, J. E. Pask {\it et al} \cite{Pask} report
a similar calculation, based on the finite element method but without
multigrid acceleration, for a supercell containing 4096 Cu atoms.
The result converged within 1 ps took  ' just 14.3 hr ' of CPU time.

\begin{figure}[h]
\centerline{\resizebox{0.8\columnwidth}{!}{\includegraphics[angle=90]{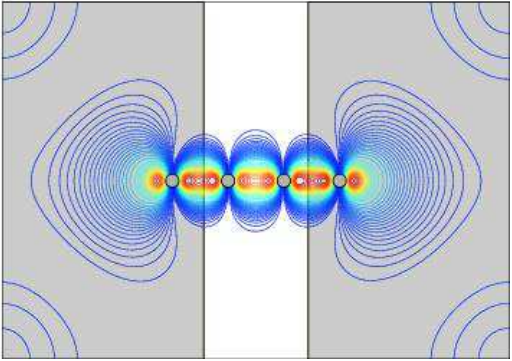}}}
\vspace{0.5cm}
\caption{Electronic structure of a four-carbon-atom chain between two jellium leads 
determined by the MIKA code. The figure shows an occupied sigma-type state localized at the atom chain.
Periodic boundary conditions within the cylidrical symmetry
have been employed along the (vertical) symmetry axis and the (horizontal) radial directions. 
The centres of pseudo ions are denoted by grey circles and the positive jellium background by 
grey shadowing. Blue contours denote low electron densities whereas the density increases towards 
yellow and red.}
\label{fig:cylinder}
\end{figure}

We have also applied the MIKA scheme in two-dimensional problems for quantum dots  
employing the current-spin-density functional theory (CSDFT), see Ref. \cite{Henri}.
Moreover, we have implemented the RQMG-method in cylindrical coordinates
enabling very efficient and accurate calculations for atomic chains,
or systems which can be described using axisymmetric jellium models.
Fig. \ref{fig:cylinder} shows a selected wavefunction of a system
where a chain of four carbon atoms is sandwhiched between two planar 
jellium leads.

\section{Summary and outlook}
\label{sec:conclusions}

In the MIKA program package the RQMG method introduced by Mandel and
McCormick \cite{McCormick} is generalized for the simultaneous solution of a desired
number of lowest eigenenergy states. The approach can be viewed to belong
to a third group of multigrid methods, in addition to FAS and
techniques where the eigenproblem is linearized.
In principle, one can use arbitrarily coarse grids in RQMG, whereas in
the other multigrid methods one has to be able to represent all the
states also on the coarsest grid.

We are convinced that our method will compete with the standard
plane-wave methods for electronic structure calculations.
However, some  straightforward programming is still required.
Implementation of the Hellmann-Feynman forces, required for the optimization 
of the ionic structures is under way.

During the RQMG V-cycle, the states are all relaxed
simultaneously and independently of each other. A parallelization
over states would therefore be natural to implement
on a shared memory architecture.
We have parallelized the MIKA codes over k-points, and over real-space domains.
The domain decomposition is the appropriate method for distributed memory
parallel computers.

\acknowledgments

We acknowledge the contributions by Henri Saarikoski, Paula Havu, Esa R\"as\"anen, 
Tero Hakala, and Sampsa Riikonen in sharing their experience of the use of 
the MIKA package in different applications and preparing the figures 
\ref{fig:positrons} (T.H.) and \ref{fig:cylinder} (P.H.).
T.T. acknowledges financial support by the Vilho, Yrj\"o and Kalle
V\"ais\"al\"a foxundation. This research has been supported by the Academy of Finland
through its Centre of Excellence Programme (2000 - 2005).


\end{multicols}

\newpage

\end{document}